\documentclass[twocolumn,showpacs,amssymb]{revtex4}
\usepackage{graphicx}
\usepackage{dcolumn}
\usepackage{bm}

\def\beq{\begin{equation}}
\def\enq{\end{equation}}
\def\ba{\begin{eqnarray}}
\def\ea{\end{eqnarray}}

\def\<{<\!\!}
\def\>{\!\!>}

\begin{document}
\input{epsf}

\title{Probing the tidal disruption flares of massive black holes with high-energy neutrinos }

\author{Xiang-Yu Wang$^{1,2}$, Ruo-Yu Liu$^{1,2}$, Zi-Gao Dai$^{1,2}$ and K. S. Cheng$^{3}$ }

\affiliation{$^1$Department of Astronomy, Nanjing University, Nanjing 210093, China\\
$^2$Key laboratory of Modern Astronomy and Astrophysics (Nanjing
University), Ministry of Education, Nanjing 210093, China\\
$^3$Department of Physics, University of Hong Kong, Hong Kong,
China.}

\begin{abstract}
The recently discovered high-energy transient  Swift
J164449.3+573451 (Sw J1644+57) is  thought to arise from the tidal
disruption of a passing star by a dormant massive black hole.
Modeling of the broadband emission suggests the presence of a
powerful relativistic jet, which contributes dominantly to the
observed X-ray emission. Here we suggest that protons can be
accelerated to ultra-high energies by internal shocks occurring in
the jets, but their flux is insufficient to account for the observed
flux of ultra-high energy cosmic rays. High energy protons can
produce $\sim0.1$-10 PeV neutrinos through photomeson interactions
with X-ray photons. The large X-ray fluence ($7\times10^{-4}{\rm
erg\, cm^{-2}}$) and high photopion efficiency, together with the
insignificant cooling of secondary mesons, result in bright neutrino
emission expected from Sw J1644+57 if the jet composition is
matter-dominated. One to several neutrinos may be detected by a
Km$^3$-scale  detector from one tidal disruption event similar to Sw
J1644+57, thereby providing a powerful probe of the composition of
the jets.

\end{abstract}

\maketitle

Massive black holes are believed to reside at the centers of most
galaxies and the vast majority of them are considered to be dormant.
It was long predicted that, if a star of mass $M_\star$ and radius
$R_\star$ passes occasionally within the disruption radius
$r_T\simeq R_\star (M_{\rm BH}/M_\star)^{1/3}$  of the dormant black
hole (where $M_{\rm BH}$ is the black hole mass), the star will be
torn apart by gravitational tidal forces, leading to a transient
accretion disk and a bright panchromatic flare\cite{TDF}. Candidates
of such tidal disruption flares (TDFs) have been
suggested\cite{TDF-candidates}. Relativistic jets are expected to
form in such accretion disk system and may produce observational
phenomena as well\cite{jets-cheng}.

The high-energy transient Swift J164449.3+573451 (hereafter Sw
J1644+57) was discovered by the Swift satellite at 2011 March 28
12:57:45 UT\cite{Burrows2011}. The facts that it occurred near the
center of a galaxy, no archival X-ray emission before detection and
its analogy with blazars in the X-ray emission suggest that it is
mostly likely to be a TDF event\cite{Bloom2011,Levan2011,other}. At
redshift $z= 0.3534$ (corresponding to a luminosity distance of
$d_L=1.8{\rm Gpc}$)\cite{Levan2011}, the early X-ray flare emission
reached a peak luminosity of $L_{\rm X}\simeq 4.3\times 10^{48} {\rm
erg \, s^{-1}}$ (isotropic equivalent) and then transits to a low
state with a median luminosity of $L_{\rm X}\simeq 2.96\times10^{47}
{\rm erg \,s^{-1}}$ in 0.4-13.5 keV over a time $\Delta T\sim10^6$ s
\cite{Burrows2011}. Correcting for the live-time fraction of the
observation, the total unabsorbed fluence is $S_{\rm
X}=7.1\times10^{-4}{\rm  erg \, cm^{-2}}$ in the observed 0.3-10.0
keV band\cite{Burrows2011}. This gives an estimate of the total
isotropic equivalent energy of $E_{\rm X}=3\times10^{53}$ erg in the
0.4-13.5 keV rest frame energy band. The observed minimum X-ray
variability time of Sw J1644+57, $t_v\simeq 100$
s\cite{Burrows2011}, constrains the size of the black hole under the
assumption that the size of the central engine determines the
shortest variability and suggest an upper limit on the massive black
hole mass $M_{\rm BH}\le 8\times 10^6 M_\odot$. The observed peak
luminosity is super-Eddington and requires a strongly anisotropic
radiation pattern with a relativistic jet of a bulk Lorentz factor
of $\Gamma\simeq 10$ pointed towards us
\cite{Burrows2011,Bloom2011}. Modeling of the broadband spectral
energy distribution (SED) also requires a powerful relativistic jet
with $\Gamma\simeq 10$, which produce dominantly the observed X-ray
emission\cite{Burrows2011,Bloom2011}. A relativistic jet is also
required to explain the radio transient\cite{Zauderer2011}. However,
how the jet is launched is not well-understood. It is believed that
the composition of the jet, whether it is matter-dominated or
magnetic field-dominated, is crucial to unveiling   the formation
mechanism of the jets.  In this paper, we suggest that jets in TDFs
can produce bright emission in high-energy neutrinos if the jet is
matter-dominated, and thus neutrino observations provide an
important tool to diagnose the jet composition.

{\em Internal shocks and particle acceleration. ---}   We consider
that a TDF event produces a relativistic matter-dominated wind of
luminosity $L_w\sim 10^{49} {\rm erg \, s^{-1}}$, moving with a bulk
Lorentz factor $\Gamma\sim 10$. Variability of the source on time
scale $t_v$, resulting in fluctuations in the wind bulk Lorentz
factor $\Gamma$ on similar time scale,  would lead to
semi-relativistic internal shocks\cite{internal-shock} in the ejecta
at a radius
\begin{equation}
R\simeq 2\Gamma^2 c t_v=6\times10^{14} \Gamma_1^2 t_{v,2} {\rm cm},
\end{equation}
which is well above the photosphere radius at $R_{ph}=\sigma_{\rm
T}L_w/(4\pi\Gamma^3m_p c^3)=1.2\times 10^{13} L_{w,49} \Gamma_1^{-3}
{\rm cm}$, where $\sigma_{\rm T}$ is the Thomson cross section. We
use c.g.s. units and the denotation $Q=10^xQ_x$ throughout the
paper.

Denoting $\epsilon_B$ as the fraction of  the wind kinetic energy
converted into magnetic fields, we have a magnetic field
$B'=(8\pi\epsilon_BL_w/4\pi R^2\Gamma^2 c)^{1/2}=1.3\times10^3
\epsilon_{B,-1}^{1/2}L_{w,49}^{1/2} R_{14.8}^{-1} \Gamma_1^{-1} {\rm
G}$, where the prime symbol represents quantities measured in the
comoving frame of the shock. It is assumed that internal shocks
accelerate protons with a spectrum $dn/d\varepsilon_p\sim
\varepsilon_p^{-2}$, where $\varepsilon_p$ is the proton energy in
the observer frame. The maximum proton energy is set by comparing
the acceleration time $t'_{\rm acc}=\alpha \varepsilon_p/(e \Gamma
B' c)=860 \alpha (\frac{\varepsilon_p}{10^{20} {\rm eV}})
\epsilon_{B,-1}^{-1/2}L_{w,49}^{-1/2} R_{14.8}\,{\rm s}$ with the
shock dynamic time $t'_{\rm dyn}=R/\Gamma c=10^3 \Gamma_1
t_{v,2}{\rm s}$, where $\alpha\sim$ a few, describing the ratio
between the acceleration time and Larmor time. This gives a maximum
proton energy of $\varepsilon_{\rm max,dyn}=
2.4\times10^{20}\alpha^{-1}
\epsilon_{B,-1}^{1/2}L_{w,49}^{1/2}\Gamma_1^{-1} {\rm eV}$. The
maximum energy is also limited by the cooling time of protons. The
synchrotron cooling time is  $t'_{syn}=6\pi m_p^4
c^3\Gamma/(\sigma_T m_e^2 \varepsilon_p B^2)=240
\epsilon_{B,-1}^{-1}L_{w,49}^{-1} R_{14.8}^2\Gamma_1^3
(\frac{\varepsilon_p}{10^{20} {\rm eV}})^{-1} {\rm s}$, which gives
a maximum proton energy $\varepsilon_{\rm max,syn}= 0.5\times10^{20}
\alpha^{-1/2}
\epsilon_{B,-1}^{-1/4}L_{w,49}^{-1/4}t_{v,2}^{1/2}\Gamma_1^{5/2}
{\rm eV}$. Another process that may prohibit the acceleration of
protons to ultra-high (UHE) energies is the photopion cooling loss.
UHE protons of energy $\varepsilon_{p}$ interact with soft photons
with energy $\epsilon_\gamma=0.15 {\rm
GeV}^2\Gamma^2/\varepsilon_p=0.15 \Gamma_1^2
\varepsilon_{p,20}^{-1}{\rm eV}$, which locate at near infrared
(NIR) band for typical jet parameters.  The number density of NIR
photons in the comoving frame is $n'_{\rm NIR}=L_{\rm NIR}/(4\pi R^2
\Gamma c \epsilon_\gamma)=2.5\times10^{14}L_{\rm
NIR,44}\Gamma_1^{-5} t_{v,2}^{-2}(\epsilon_\gamma/0.15{\rm
eV})^{-1}{\rm cm^{-3}}$, where $L_{\rm NIR}\simeq10^{44} {\rm erg
s^{-1}}$ is the NIR luminosity at times a few days after the initial
trigger \cite{Burrows2011}.  The photopion cooling time in the
comoving frame is $t'_{p\gamma}=1/(\sigma_{p\gamma}n'_{\rm NIR}c
K_{p\gamma})=1200L_{\rm NIR,44}^{-1}\Gamma_1^5
t_{v,2}^2(\epsilon_\gamma/0.15{\rm eV}){\rm s}$, where $K_{p\gamma}$
is the inelasticity  and $\sigma_{p\gamma}=5\times 10^{-28} {\rm
cm^{-2}}$ is the peak cross section at the $\Delta$ resonance. By
equating $t'_{acc}$ with $t'_{p\gamma}$, we get the maximum proton
energy limited by the photopion cooling process, $\varepsilon_{\rm
max,p\gamma}=1.3\times10^{20}\alpha^{-1}
\epsilon_{B,-1}^{1/4}L_{w,49}^{1/4}L_{\rm
NIR,44}^{-1/2}\Gamma_1^{5/2}t_{v,2}^{1/2} {\rm eV}$.  Thus, internal
shocks in TDFs can accelerate protons to energies above $10^{19}$
eV, in support of the earlier suggestion that TDFs can produce
ultra-high energy cosmic rays (UHECRs)\cite{Farrar2009}. However,
the flux of such UHE protons contributed by TDFs in the universe is
insufficient to explain the observed flux of UHECRs, as we show
below.

{\em UHECR flux---}The Swift BAT, with a field of view of $S_{\rm
FOV}\simeq4\pi/7$ sr, has detected one such event in a time $T\simeq
7 {\rm years}$ at a flux that would have been detectable up to  a
luminosity distance of $d_{max}=5{\rm Gpc}$ \cite{Burrows2011}.
Therefore we will assume that the rate of Sw J1644+57-like event is
\begin{equation}
\dot R=\frac{4\pi}{S_{\rm FOV}} \frac{1}{T} \frac{1}{(4/3)\pi
d_{max}^3}\simeq2\times10^{-12} {\rm Mpc^{-3} yr^{-1}}
\end{equation}
and the energy injection rate in X-rays is $\dot\varepsilon_{\rm
X}=\dot R E_{\rm X}=6\times10^{41} {\rm erg Mpc^{-3} yr^{-1}}$.
Following ref.\cite{Murase2006}, the total energy in accelerated
protons $E_p$ can be parameterized by $E_p=\xi_{\rm p} E_{\rm X}$,
where the non-thermal baryon loading factor $\xi_{\rm p}$ can be
expressed by $\xi_{\rm
p}=10\varsigma_p\eta_{e}^{-1}(0.1/\epsilon_e)$, $\epsilon_e$ is the
fraction of the shock internal energy that goes into non-thermal
electrons, $\eta_{e}$ is the radiative efficiency of these
electrons, and $\varsigma_p$ is the proton acceleration efficiency.
Modeling of the afterglows of gamma-ray bursts gives a typical value
$\epsilon_e\simeq0.1$ for relativistic shocks\cite{Kumar}, so the
typical value of $\xi_{\rm p}$ would be $\sim10$. Thus the
differential energy injection rate in protons is
\begin{equation}
\varepsilon_{p}^2 \frac{d\dot n}{d\varepsilon_{p}}=\frac{\xi_{\rm
p}\dot R E_{\rm X}}{{\rm
ln}(\varepsilon_{p,max}/\varepsilon_{p,min})}\simeq 6\times10^{41}
\xi_{\rm p,1} {\rm erg Mpc^{-3} yr^{-1}},
\end{equation}
where $\varepsilon_{p,max}$ and $\varepsilon_{p,min}$ are,
respectively, the maximum and minimum energy of accelerated protons
and we have used ${\rm
ln}(\varepsilon_{p,max}/\varepsilon_{p,min})=10$ in the last step.
For $\xi_{\rm p}$ of the order $\sim 10$, this rate is much  smaller
than the required energy generation rate of cosmic rays   per energy
decade from $0.5-20 \times10^{44} {\rm erg Mpc^{-3} yr^{-1}}$
deduced by different authors\cite{Katz2009}. Note, however, that it
was only the presence of short, powerful bursts early on that
alerted us to its presence, so we can not exclude  the possibility
that many other similar, but rather less variable, events could  be
undetected.

{\em Pion production. ---} Now we consider the neutrino emission
produced by these protons interacting with the soft  photons in the
sources. Protons lose $\sim20\%$ of their energy at each $p\gamma$
interaction, dominated by the $\Delta$ resonance.  Approximately
half of the pions are charged and decay into high energy neutrinos
$\pi^+\rightarrow\mu^+ + \nu_\mu \rightarrow e^+ + \nu_e +
\bar\nu_\mu+\nu_\mu$, with the energy distributed roughly equally
among the decay products.   The fraction of energy lost by protons
to pions is $f_{p\gamma}=R/(\Gamma c t'_{p\gamma})$. Denoting by
$n(\epsilon_\gamma)d\epsilon_\gamma$ the number density of photons
in the energy range $\epsilon_\gamma$ to
$\epsilon_\gamma+d\epsilon_\gamma$, the cooling time of protons in
the shock comoving frame for $p\gamma$   process is given by
\begin{equation}
t'_{p\gamma}=\frac{c}{2\Gamma_p^2} \int_{\epsilon_{th}}^\infty
d\epsilon \sigma (\epsilon)
K(\epsilon)\epsilon\int_{\epsilon/2\Gamma_p}^{\infty} dx x^{-2}
n(x),
\end{equation}
where $\Gamma_p=\varepsilon_p/\Gamma m_p c^2$, $\sigma$ and $K$ are
respectively the cross section and the inelasticity for $p\gamma$
process \cite{Stecker1968}. The spectrum from infrared to X-ray
frequencies of Sw J1644+57 can be approximately described by a
broken power-law with
$n(\epsilon_\gamma)=n_b(\epsilon_\gamma/\epsilon_b)^{-\alpha}$ for
$\epsilon_\gamma<\epsilon_b$ and
$n(\epsilon_\gamma)=n_b(\epsilon_\gamma/\epsilon_b)^{-\beta}$ for
$\epsilon_\gamma>\epsilon_b$, where $\epsilon_b\sim 1{\rm KeV}$,
$\alpha\simeq2/3$ and $\beta\simeq 2$\cite{Burrows2011}.
Approximating the integral by the contribution from the resonance we
obtain
\begin{equation}
f_{p\gamma}(\varepsilon_p)\simeq0.35\frac{L_{\rm X,47.5}}{\Gamma_1^4
t_{v,2}\epsilon_{\rm b, KeV}} \left \{
\begin{array}{ll}
(\varepsilon_p/\varepsilon_{pb})^{\beta-1}
(\varepsilon_p<\varepsilon_{pb})\\
(\varepsilon_p/\varepsilon_{pb})^{\alpha-1}
(\varepsilon_p>\varepsilon_{pb})
\end{array} \right .
\end{equation}
where $\varepsilon_{pb}=0.15 {\rm
GeV^2}\Gamma^2/\epsilon_b=1.5\times10^{16}\Gamma_1^2\epsilon_{\rm
b,KeV}^{-1} {\rm eV}$ is the proton break energy.  To include the
effect of multi-pion production  and high inelasticity (which
increases from $\simeq0.2$ at energies not far above the threshold
to $\sim 0.5-0.6$ at  energies where multi-pion production
dominates) at high energies\cite{Mucke1999,Atoyan2003}, the above
estimate of $f_{p\gamma}$ should be multiplied by factor of $\sim
2$. As the neutrino energy is $\sim5\%$ of the proton energy, the
neutrino flux will peak at $\varepsilon_{\nu b}\simeq
7.5\times10^{14}\Gamma_1^2\epsilon_{\rm b,KeV}^{-1} {\rm eV}$.

In the modeling of the SED of Sw J1644+57,  upper limits from Fermi
and VERITAS require $\Gamma<20$ in the X-ray emitting region
\cite{Burrows2011}. It is useful to express $f_{p\gamma}$ as a
function of the pair production optical depth $\tau_{\gamma\gamma}$.
The optical depth for pair production of a photon of energy
$\varepsilon_h$ is $\tau_{\gamma\gamma}(\epsilon_h)=\frac{R}{\Gamma
l_{\gamma\gamma}}=\frac{R}{ \Gamma}\frac{\sigma_T}{16}
\frac{U_\gamma \epsilon_h}{\Gamma (m_e c^2)^2}$, where
$l_{\gamma\gamma}$ is the mean free path. The fraction of energy
lost by protons to pions is
$f_{p\gamma}\simeq\frac{R}{\Gamma}\frac{U_\gamma}{2\varepsilon'_p}
 \sigma_{p\gamma} K_{p\gamma} (\varepsilon_p/\varepsilon_{pb})^{\alpha-1}$ for
protons with energy $\varepsilon_p>\varepsilon_{pb}$. Thus, we
have\cite{Waxman-Bahcall97}
\begin{equation}
f_{p\gamma}(\varepsilon_{p})\simeq0.5\tau_{\gamma\gamma}(100{\rm
MeV})\left(\frac{\varepsilon_{pb}}{1.5\times10^{16}{\rm
eV}}\right)\left(\frac{\varepsilon_{p}}{\varepsilon_{pb}}
\right)^{\alpha-1} .
\end{equation}
Modeling of the SED of Sw J1644+57 requires
$\tau_{\gamma\gamma}(100{\rm MeV})>1$ \cite{Burrows2011}, so we
conclude that  a significant fraction ($>50\%$) of the energy of
protons accelerated to energies larger than the break energy,
$\varepsilon_{pb}\sim 10^{16}$eV, would be lost to pion production.

{\em Meson cooling. ---} The neutrino production efficiency will be
reduced  if the secondary mesons suffer from cooling before decaying
to neutrinos and other  products \cite{Rachen}. The pions and muons
suffer from radiative cooling due to both synchrotron emission and
inverse-Compton emission. The total radiative cooling time for pions
is $t'_{\pi, rad}={3 m_\pi^4 c^3}/[{4\sigma_T m_e^2 \epsilon'_{\pi}
U'_B(1+f_{\rm IC})}]\simeq2\times10^6 ({\epsilon'_{\pi}}/{1 {\rm
TeV}})^{-1} \epsilon_{B,-1}^{-1}L_{w,48}^{-1} R_{14.5}^2 \Gamma_1^2
\,{\rm s}$, where $U'_B$ is the energy density of the magnetic filed
in the shock region and $f_{\rm IC}\le1$ is the correction factor
accounting for the inverse-Compton loss. By comparing this cooling
time $t'_{\pi, rad}$ with the lifetime of pions
$\tau'_{\pi}=\gamma_{\pi} \tau=1.9\times10^{-4} (\epsilon'_{\pi}/1
{\rm TeV}) \,{\rm s}$ in the shock comoving frame, where
$\gamma_{\pi}$ and $\tau$ are the pion Lorentz factor and proper
lifetime, one can find  a critical energy (in the observer frame)
for pions, above which the effect of radiative cooling starts to
suppress the neutrino flux,
\begin{equation}
\varepsilon_{\pi, rad}=6.3\times10^5
\epsilon_{B,-1}^{-1/2}L_{w,49}^{-1/2}\Gamma_1^4 t_{v,2} {\rm TeV}.
\end{equation}
Similarly,  by comparing the radiative cooling time $t'_{\mu, rad}$,
with the lifetime of muons $\tau'_{\mu}$, one can obtain a critical
energy for muons, above which the effect of radiative cooling starts
to suppress the anti-muon neutrino flux,
\begin{equation}
\varepsilon_{\mu, rad}=3.2\times10^4
\epsilon_{B,-1}^{-1/2}L_{w,49}^{-1/2}\Gamma_1^4 t_{v,2} {\rm TeV}.
\end{equation}
The above estimates lead us to conclude that the neutrino flux below
$\sim10^{16}{\rm eV}$ is not affected by the meson cooling for
typical parameters of TDFs. At higher energies, however, pion
cooling and muon cooling will suppress  the neutrino  flux  by a
factor approximately given, respectively, by \cite{Razzaque}
\begin{equation}
\zeta_{\pi}={\rm Min}\{t'_{\pi, rad}/\tau'_{\pi}, 1\},
\zeta_{\mu}={\rm Min}\{t'_{\mu, rad}/\tau'_{\mu},1\}.
\end{equation}

{\em The spectrum and  flux of the neutrino flare. ---} The fluence
spectrum of the muon neutrino ($\nu_\mu+\bar\nu_\mu$) emission from
one TDF is
\begin{equation}
{\varepsilon_\nu^2}\Phi_\nu=\varepsilon_p^2\frac{dn_p}{d\varepsilon_p}\frac{f_{p\gamma}
\zeta_{\pi}(1+\zeta_{\mu})}{8}= \frac{E_{\rm X}}{32\pi d_L^2} \frac{
\xi_{\rm p}f_{p\gamma} {\zeta_{\pi}(1+\zeta_{\mu})}}{{\rm
ln}(\varepsilon_{p,max}/\varepsilon_{p,min})},
\end{equation}
where $\varepsilon_p^2\frac{dn_p}{d\varepsilon_p}=E_p/[4\pi
d_L^2{\rm ln}(\varepsilon_{p,max}/\varepsilon_{p,min})]$ is the
differential proton fluence produced by one TDF of total energy
$E_p=\xi_{\rm p}E_{\rm X}$ in protons and the factor 1/8 represents
that the neutrinos produced by pion decay carry one-eighth of the
energy lost by protons to pion production, since charged and neutral
pions are produced with roughly equal probability and muon neutrinos
carry roughly one-fourth of the pion energy in pion decay. Fig.1
shows the expected muon neutrino fluence spectra from Sw J1644+57,
obtained by using the Lorentzian form for the photopion production
cross section at the resonance peak plus a component contributed by
multi-pion production at higher energies \cite{Mucke2000} in
calculating $t'_{p\gamma}$ with Eq.(4). The initial rise in the
spectrum at low energies is caused by the increasing pion production
efficiency with energy, while the mild steepening and sharp
steepening seen at higher energies are caused by muon cooling and
pion cooling respectively.
\begin{figure}
\centerline {\epsfxsize=3.5in \epsfbox{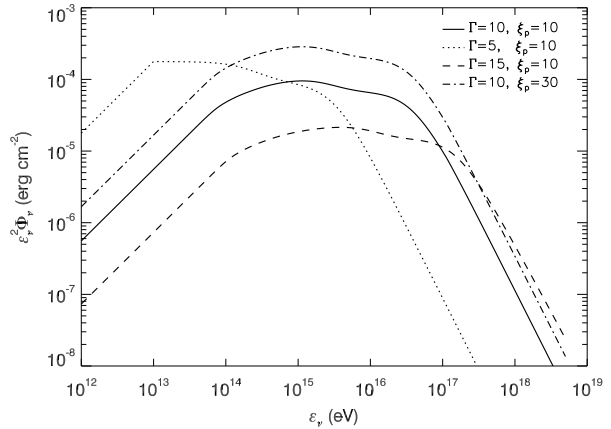}} \caption{The
expected muon neutrino ($\nu_\mu$+$\bar\nu_\mu$) spectra from Sw
J1644+57 for different jet parameters. In all lines, $S_{\rm
X}=7\times10^{-4}{\rm erg cm^{-2}}$, $L_{\rm X}=3\times10^{47}{\rm
erg s^{-1}}$, $t_v=100 {\rm s}$, $\epsilon_{b}=1{\rm KeV}$ and
$\epsilon_B=0.1$ are used. }
 \label{fig:spectrum}
\end{figure}

Now we estimate the number of neutrinos that can be detected from
one TDF event similar to Sw J1644+57. The detection efficiency in
water or ice of ultra-relativistic upward-going muon neutrinos with
energies $\varepsilon_\nu$  is $P_{\nu\mu}\simeq
7\times10^{-5}(\varepsilon_\nu/10^{4.5}{\rm GeV})^\kappa$,  where
$\kappa=1.35$ for $\varepsilon_{\nu}< 10^{4.5}{\rm GeV}$, and
$\kappa = 0.55$ for $\varepsilon_{\nu}
> 10^{4.5}{\rm GeV}$\cite{probablity}.  For  neutrino fluence spectrum parameterized by
${\varepsilon_\nu^2}\Phi_\nu$, the number of $\nu_\mu$ and
$\bar\nu_\mu$ above a certain energy $\varepsilon_{\nu 0}$ detected
by a km$^3$-scale neutrino detector, such as IceCube, with area
$A=10^{10}A_{10}{\rm cm^2}$ is
\begin{equation}
\begin{array}{ll}
N_\nu(>\varepsilon_{\nu 0}=10^{4.5}{ \rm
GeV})=\int_{\varepsilon_{\nu 0}}^{\infty}\Phi_\nu P_{\nu\mu} A
d\varepsilon_\nu \\\simeq2 \xi_{{\rm
p},1}A_{10}{f_{p\gamma}(\varepsilon_{p
b})}\left(\frac{E_X}{3\times10^{53}{\rm
erg}}\right)\left(\frac{d_L}{1.8{\rm Gpc}}\right)^{-2},
\end{array}
\end{equation}
where we have used ${\rm
ln}(\varepsilon_{p,max}/\varepsilon_{p,min})=10$, $\zeta_{\pi}\simeq
1$, $\zeta_{\mu}\simeq 1$, $\alpha=2/3$ and $\beta=2$ in the last
step. As $f_{p\gamma}>0.5$, we expect $\ge 1$ neutrinos  above 30
TeV could be detected from Sw 1644+57 by Km$^3$-scale detectors for
$\xi_{\rm p}=10$. A careful calculation of the number of neutrinos
above a certain energy as a function of the neutrino energy is shown
in Fig. 2. A lower bulk Lorentz factor or a higher wind luminosity
(i.e. a larger $\xi_{\rm p}$)  leads to a larger number of neutrinos
that can be detected. If a similar event to Sw J1644+57 occurs at a
closer distance (e.g. at $z=0.2$) in future,  more neutrinos would
be detected as well.

\begin{figure}
\centerline {\epsfxsize=3.5in \epsfbox{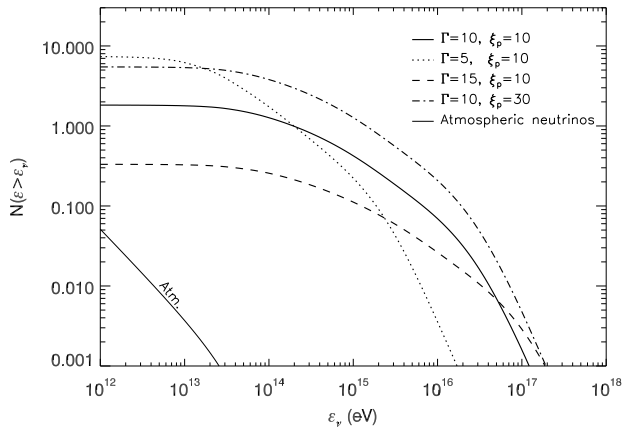}} \caption{The
expected number of neutrinos ($\nu_\mu$+$\bar\nu_\mu$) above a
certain energy detected from Sw J1644+57 by Km$^3$-scale neutrino
detectors such as ICecube. The thin solid line represent the
background atmospheric neutrinos. The same parameters as in Fig.1
are used. }
 \label{fig:spectrum}
\end{figure}

Neutrino detection from  TDFs can be assured only if the number of
background counts is smaller than one. The number of atmospheric
neutrinos above 30 TeV expected in the direction of the source
during the flare period is
\begin{equation}
\begin{array}{ll}
N_{atm}(>10^{4.5}{ \rm GeV})=\int_{10^{4.5}{ \rm GeV}}^{\infty}d
\varepsilon_\nu\int d\Omega \int d t
\frac{F_\nu^{atm}}{\varepsilon_\nu^2}P_{\nu\mu} A \\
\simeq 3\times10^{-3} \left(\frac{\Delta T}{10^6{\rm s}}\right)
A_{10} \left(\frac{\theta}{1^\circ}\right)^2,
\end{array}
\end{equation}
where $\Delta T$ is characteristic duration of the TDF,
$\theta\simeq0.5^\circ-0.6^\circ$ is the angular resolution of the
neutrino detector at 30TeV-PeV\cite{Ahrens2004}, $F_\nu^{atm}$ is
the cosmic-ray induced atmospheric neutrino background flux. We fit
the atmospheric neutrino  data measured by Icecube \cite{Abbasi2011}
with a single power-law function, which gives $F_\nu^{atm}\simeq
4.7\times 10^{-8}{\rm erg cm^{-2} s^{-1} sr^{-1}}
(\varepsilon_\nu/1{\rm TeV})^{-\beta}$ with $\beta\simeq1.44$ in the
energy range of 0.1-400 TeV. Since the number of atmospheric
neutrinos above 30 TeV expected in the direction of the source
during the flare period is much smaller than one,  a detection of
two neutrinos at energies above 30 TeV from TDF sources will be
highly significant.

{\em Discussions. ---} Neutrino emission has been predicted to be
produced by relativistic jets in gamma-ray bursts
\cite{Waxman-Bahcall97,GRB-neutrino, Murase2006}, active galactic
nuclei \cite{AGN-neutrino} and
microquasars\cite{microquasar-neutrino}. Observations of Sw 1644+57
suggest that powerful jets are formed in TDFs, which have larger
fluence in X/$\gamma$-rays than the brightest GRBs and have higher
X-ray luminosity than AGNs. There are three factors that are
favorable for bright neutrino emission produced in such TDFs: 1)
large fluence in the X-ray emission, which suggests large fluence in
accelerated protons; 2) high pion production efficiency as implied
by the high opacity of high-energy gamma-rays, which leads to a high
fraction of the proton energy lost into secondary pions; 3)
insignificant radiative cooling of secondary pions and muons, which
leads to an almost flat neutrino spectrum up to  $\sim 10^{16}$ eV.
The main uncertainty lies in the ratio between the energy density of
protons and the energy density in X-rays. In the magnetic
field-dominated jet model, the proton energy density is subdominant,
so the neutrino flux would be low, whereas in the matter-dominated
jet model, we expect one to several neutrinos detectable by
Km$^3$-scale neutrino detectors from TDFs similar to Sw 1644+57. The
Swift BAT, with a field of view of 4$\pi$/7 sr, has detected one TDF
in 7 years, so the all-sky rate of TDFs  would be one event in every
one year. For neutrino detectors such as Icecube that has a 2$\pi$
sr field of view, we expect one TDF event in the field of view of
Icecube every two years, if the electro-magnetic counterparts can be
identified. Therefore neutrino observations provide a promising
approach to diagnose the composition of the jets resulted from tidal
disruption of stars by massive black holes in the galaxy center.

This work is supported by the NSFC under grants 10973008, 10873009
and 11033002, the 973 program under grants 2009CB824800 and
2007CB815404,  the Program for New Century Excellent Talents in
University, the Qing Lan Project and the Fok Ying Tung Education
Foundation. KSC is supported by a GRF Grant of the Government of the
Hong Kong SAR under HKU 7011/10P.

\end{document}